\documentclass{aip-cp}

\usepackage[numbers]{natbib}
\usepackage{rotating}
\usepackage{graphicx}

\begin{document}

\title{$^{100}$Mo-enriched Li$_2$MoO$_4$ scintillating bolometers for $0\nu 2\beta$ decay search:
from LUMINEU to CUPID-0/Mo projects}

\author[]{D.V.~Poda for LUMINEU, EDELWEISS, and CUPID-0/Mo Collaborations}

\affil[]{CSNSM, Univ. Paris-Sud, CNRS/IN2P3, Universit\'e Paris-Saclay, 91405 Orsay, France}
\affil[]{Institute for Nuclear Research, 03028 Kyiv, Ukraine}

\maketitle

\begin{abstract}
A scintillating bolometer technology based on $^{100}$Mo-enriched lithium molybdate (Li$_2$$^{100}$MoO$_4$) 
crystals has been developed by LUMINEU to search for neutrinoless double-beta ($0\nu 2\beta$) decay of $^{100}$Mo. 
The results of several low temperature tests at underground environments have proved the reproducibility 
of high detector performance and crystal radiopurity: in particular $\sim$5--6~keV FWHM energy 
resolution and at least 9$\sigma$ rejection of $\alpha$'s in the vicinity of the $0\nu 2\beta$ decay of $^{100}$Mo 
(3034 keV) and below 10~$\mu$Bq/kg bulk activity of $^{228}$Th and $^{226}$Ra. A modest acquired exposure 
(0.1~kg$\times$yr) is a limiting factor of the LUMINEU experiment sensitivity to the $0\nu 2\beta$ decay 
half-life of $^{100}$Mo ($T_{1/2}$ $\geq$ 0.7$\times$10$^{23}$ yr at 90\% C.L.), however the two-neutrino 
$2\beta$ decay has been measured with the best up to-date accuracy, 
$T_{1/2}$ = $\left[6.92 \pm 0.06(\mathrm{stat.}) \pm 0.36(\mathrm{syst.})\right] \times 10^{18}$ yr.
The applicability of the LUMINEU technology for a tonne-scale $0\nu 2\beta$ decay bolometric project CUPID 
is going to be demonstrated by the CUPID-0/Mo experiment with $\sim$5~kg of $^{100}$Mo embedded in forty 
0.2~kg Li$_2$$^{100}$MoO$_4$ scintillating bolometers. A first phase of the experiment with twenty 
Li$_2$$^{100}$MoO$_4$ detectors is in preparation at the Modane underground laboratory (France) to start 
by the end of 2017.
\end{abstract}

\section{INTRODUCTION}
Searches for neutrinoless double-beta decay ($0\nu 2\beta$) --- a lepton number violating 
spontaneous nuclear transition which requires a Majorana nature of neutrinos (e.g. see the recent 
review \cite{Vergados:2017} and references herein) --- are among the hottest worldwide experimental efforts 
in Astroparticle physics. The sensitivity to the $0\nu 2\beta$ half-lives already / to be achieved by the present 
generation leading $0\nu 2\beta$ experiments (lim$T_{1/2}^{0\nu 2\beta} \sim 10^{24} - 10^{26}$~yr 
\cite{Vergados:2017,Gando:2017}) has to be improved by two orders of magnitude to make a further significant 
progress in the field \cite{Bilenky:2015}. Such enhancement would be possible with a new / advanced technology, 
which can provide near zero-background conditions in the region of interest (around the $Q$-value of 
the $0\nu 2\beta$ transition, $Q_{\beta \beta}$) for a ton-scale detector over years of exposure. 
LUMINEU (Luminescent Underground Molybdenum Investigation for NEUtrino mass and nature) is a French-funded 
project (2012--2017) aiming at the development of such challenging technology based on $^{100}$Mo-enriched 
zinc and lithium molybdate (ZnMoO$_4$ and Li$_2$MoO$_4$) scintillating bolometers. Since recently, 
LUMINEU is part of an R\&D activity towards CUPID (CUORE Upgrade with Particle ID) \cite{CUPID,CUPID_RD}, 
a next-generation $0\nu 2\beta$ project aiming at using as much as possible the infrastructure of the present 
ton-scale bolometric experiment CUORE. In this paper we will overview the main LUMINEU achievements concerning 
the R\&D of Li$_2$MoO$_4$ scintillating bolometers which led to the preparation of a $0\nu 2\beta$ experiment 
CUPID-0/Mo aiming at a demonstration of the suitability of the LUMINEU technology for CUPID.

\section{LUMINEU R\&D OF Li$_2$$^{100}$MoO$_4$ SCINTILLATING BOLOMETERS}

\begin{figure}[ht]
  \includegraphics[width=0.45\textwidth]{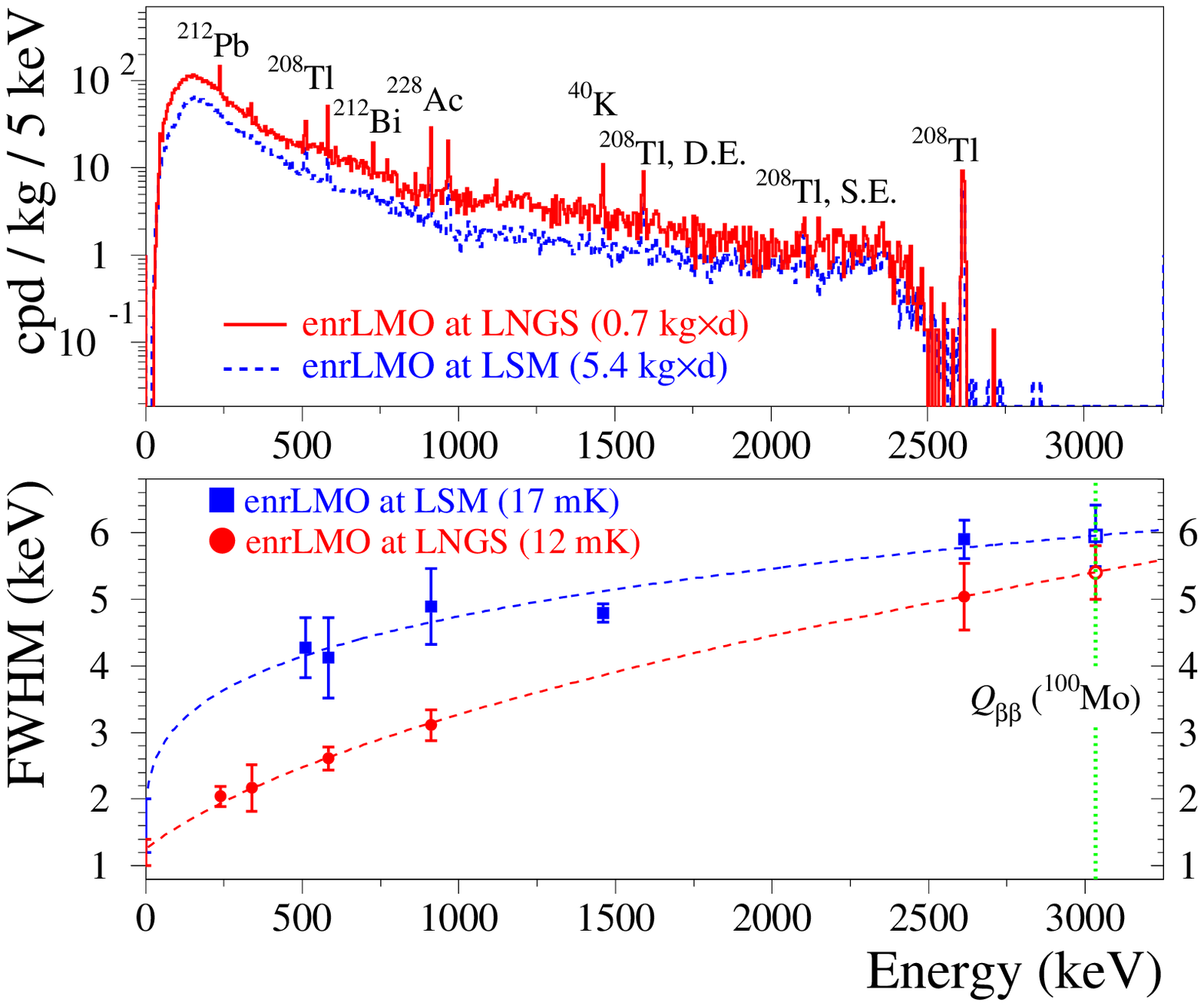}
  \includegraphics[width=0.45\textwidth]{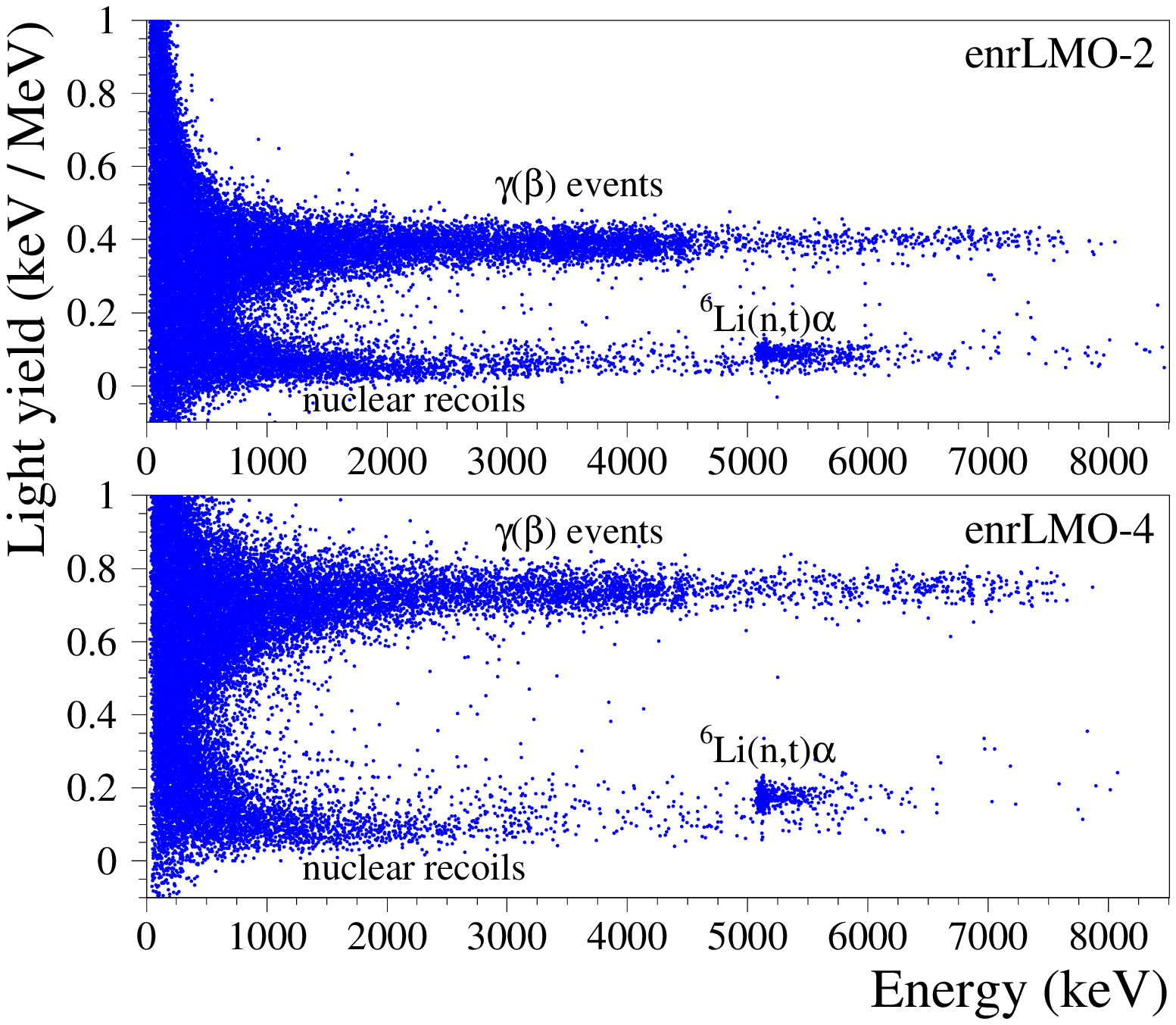}
\caption{Left panel: The energy spectra of the $^{232}$Th source and the dependence of FWHM 
energy resolution measured by a single 0.2 kg Li$_2$$^{100}$MoO$_4$ module and four-bolometer array operated 
at LNGS (12 mK; \cite{Armengaud:2017}) and LSM (17 mK), respectively. 
The expected resolution at $Q_{\beta \beta}$ of $^{100}$Mo is $\sim$5--6 keV FWHM shown as an open circle 
and an open square with error bars according to the fits to the LNGS and LSM data respectively.
Right panel: The  scatter-plots of light yield vs. heat energy of the AmBe data (290 h) of the enrLMO-2 and enrLMO-4 
detectors operated at LSM without and with a reflecting film, respectively.}
\label{fig:Calibration}
\end{figure}

An R\&D of $^{100}$Mo-enriched Li$_2$MoO$_4$ scintillating bolometers has been accomplished by LUMINEU 
as follows:
\begin{itemize}
\item Development of molybdenum purification methods \cite{Berge:2014}.
\item Screening selection of a commercial ultra-pure Li$_2$CO$_3$ powder \cite{Armengaud:2017}. 
An R\&D of Li$_2$CO$_3$ powder purification is still ongoing to ensure its high purity.
\item Optimization of the crystal growth with the help of the low-temperature-gradient 
Czochralski technique \cite{Grigoryeva:2017}. Investigation of a double crystallization to further improve 
crystal's quality and radiopurity \cite{Armengaud:2017,Grigoryeva:2017}.
\item Underground tests of Li$_2$MoO$_4$ and Li$_2$$^{100}$MoO$_4$ ($^{100}$Mo enrichment is $\sim$97\%) 
scintillating bolometers \cite{Armengaud:2017}.
\item A pilot $2\beta$ experiment with four 0.2-kg Li$_2$$^{100}$MoO$_4$ cryogenic detectors.
\end{itemize}
The results of the LUMINEU R\&D demonstrated that the crystallization technology is mature for a mass production 
of large (up to $\oslash4.5 \times 15$~cm or $\oslash6 \times 10$~cm), perfect optical quality, highly radiopure 
Li$_2$$^{100}$MoO$_4$ scintillators and that the bolometric technology is well reproducible 
in terms of high detector's performance (e.g. see in Table~\ref{tab:enrLMO} and Figure~\ref{fig:Calibration}), 
which altogether completely satisfies the LUMINEU specifications.  

\begin{table}[ht]
\caption{Performance and radiopurity of four Li$_2$$^{100}$MoO$_4$ scintillating bolometers 
tested at 17 mK in the EDELWEISS set-up at the Modane underground laboratory (LSM, France). 
The energy resolution (FWHM) is measured at 2615 keV $\gamma$ quanta of $^{208}$Tl during 
a $^{232}$Th calibration. The light yield for $\gamma$($\beta$)'s (LY$_{\gamma(\beta)}$) 
and the $\alpha$/$\gamma$($\beta$) separation efficiency are extracted from an AmBe calibration data 
(the 2.5--3.5 MeV $\gamma$($\beta$)'s and $\alpha$+t events with $\sim$5.1 MeV electron-equivalent energy 
caused by $^6$Li(n,t)$\alpha$ reaction were used). The radioactive contamination 
is estimated from the analysis of the energy spectra of $\alpha$ events accumulated in the background 
measurements. The performed analysis is similar to the one described in detail in \cite{Armengaud:2017}.}
\tabcolsep7pt\begin{tabular}{cccccccc}
\hline
\bf{Detector's}	& \bf{Crystal's}	& \bf{FWHM (keV)}		& \bf{LY$_{\gamma(\beta)}$} & \bf{$\alpha/\gamma(\beta)$ Separation}	& \multicolumn{3}{c}{\bf{Activity ($\mu$Bq/kg)}} \\
\bf{ID}					& \bf{mass (g)}		& \bf{at 2615 keV}	& \bf{(keV/MeV)} & \bf{above 2.5 MeV}		& \bf{$^{228}$Th}	& \bf{$^{226}$Ra}	& \bf{$^{210}$Po} \\
\hline
enrLMO-1	& 186	& 5.8(6)	& 0.41	& 9$\sigma$			& $\leq$4				& $\leq$6			& 450(30) \\	
enrLMO-2	& 204	& 5.7(6)	& 0.38	& 9$\sigma$			& $\leq$6				& $\leq$11		& 200(20) \\		
enrLMO-3	& 213	& 5.5(5)	& 0.73	& 14$\sigma$		& $\leq$3				& $\leq$3			& 76(10) \\	
enrLMO-4  & 207	& 5.7(6)	& 0.74	& 14$\sigma$		& $\leq$5				& $\leq$9			& 20(6) \\
\hline
\end{tabular}
\label{tab:enrLMO}
\end{table}

\section{INVESTIGATION OF $2\beta$ DECAY OF $^{100}$Mo}

\begin{figure}[ht]
  \includegraphics[width=0.45\textwidth]{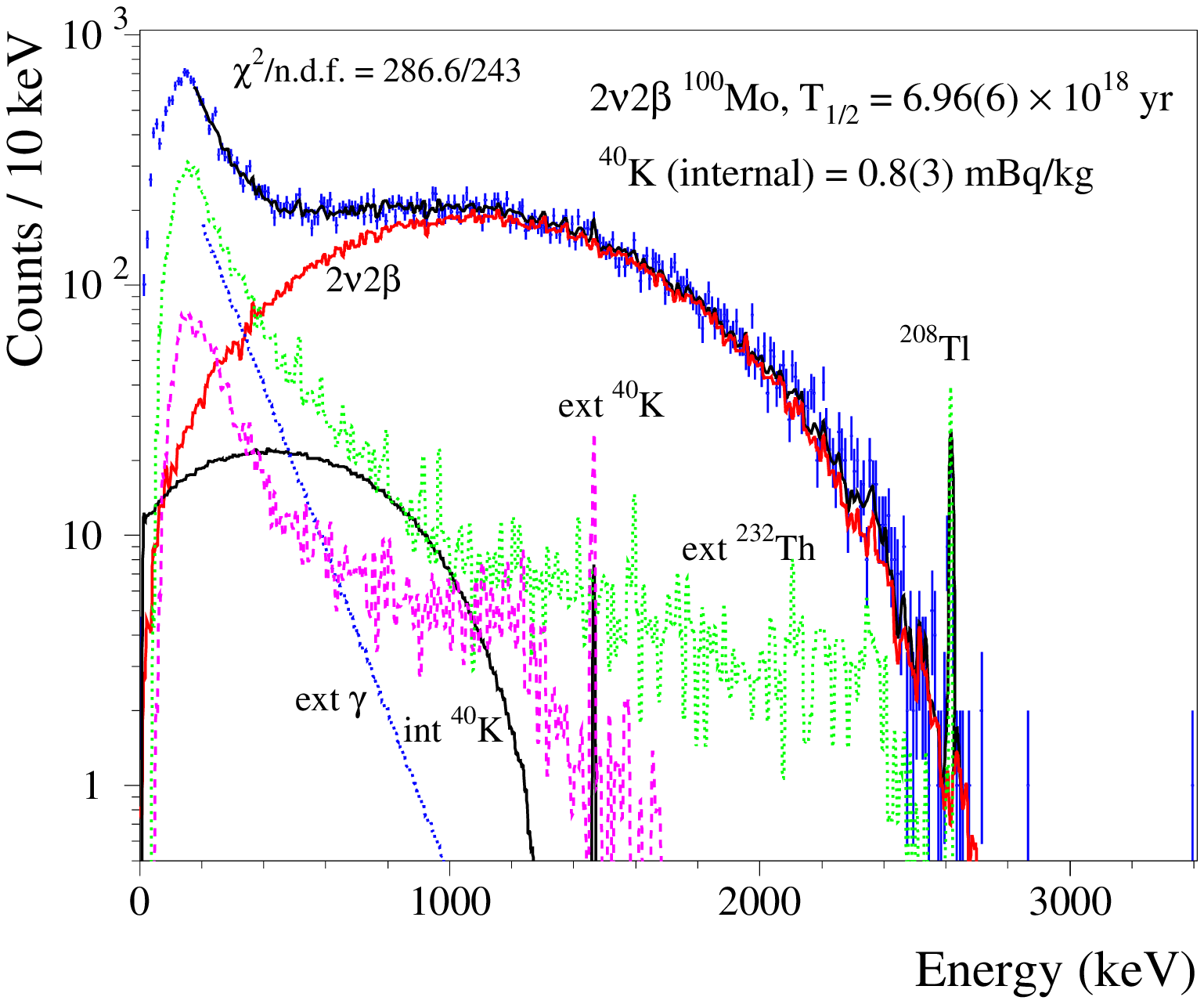}
  \includegraphics[width=0.45\textwidth]{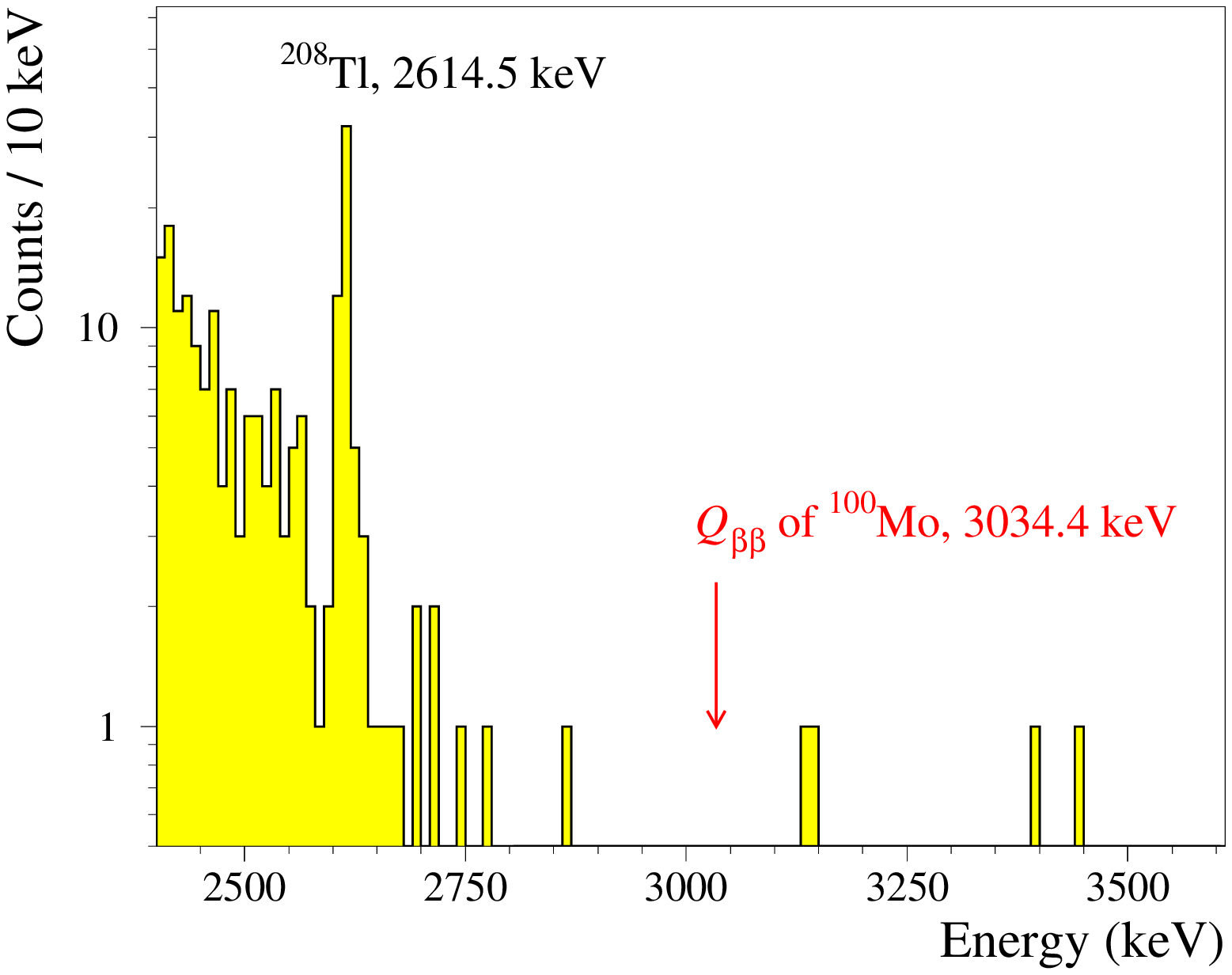}
\caption{Left panel: The background energy spectrum of $\gamma$($\beta$) events accumulated 
by an array of four 0.2~kg Li$_2$$^{100}$MoO$_4$ scintillating bolometers over 0.04~kg$\times$yr of $^{100}$Mo 
exposure in the EDELWEISS set-up. The data are fitted by a simple model constructed from a distribution 
of the $2\nu 2\beta$ decay of $^{100}$Mo ($2\nu 2\beta$), a bulk / external $^{40}$K (int / ext $^{40}$K), 
an external $^{232}$Th (ext $^{232}$Th) and an exponential function to describe other contribution of 
$\gamma$ quanta from a residual pollution of the set-up (ext $\gamma$). 
Right panel: The energy spectrum of $\gamma$($\beta$) events in the vicinity of $0\nu 2\beta$ decay 
of $^{100}$Mo extracted from the background measurements with 0.2~kg Li$_2$$^{100}$MoO$_4$ 
cryogenic detectors (0.06~kg$\times$yr of $^{100}$Mo) operated in the EDELWEISS set-up.}
\label{fig:Background}
\end{figure}

The LUMINEU pilot $2\beta$ experiment was able to perform a precise investigation of the two neutrino double-beta 
decay of $^{100}$Mo, as it is illustrated in Figure \ref{fig:Background} (left). The analysis 
is similar to the one described in \cite{Armengaud:2017}. The half-life value is derived with the best up to-date accuracy (see Table \ref{tab:DBD_100Mo}). Figure \ref{fig:Background} (right) shows a few events registered above 
the 2615~keV peak with an average rate 1.1(2) cpd/kg, but no events are observed in the 200-keV-wide energy 
interval centered at $Q_{\beta \beta}$ of $^{100}$Mo. Thus, we set a lower 
half-life limit $T_{1/2}^{0\nu 2\beta} \geq$ 0.7$\times$10$^{23}$~yr at 90\% C.L. which is about one order 
of magnitude weaker than the NEMO-3 result 
($T_{1/2}^{0\nu 2\beta} \geq 1.1 \times 10^{24}$~yr at 90\% C.L. \cite{Arnold:2015}), 
but it is achieved over an exposure of $^{100}$Mo shorter by a factor 600 (0.06 vs. 34.3 kg$\times$yr). 

\begin{table}[ht]
\caption{The most precise half-lives of $2\nu 2\beta$ decay of $^{100}$Mo (g.s.$\rightarrow$g.s.) and the signal-to-background 
ratio (S/B) achieved by means of a passive and an active $\beta \beta$ source techniques. Other results can be found in \cite{Barabash:2015}.}
\tabcolsep7pt\begin{tabular}{cccccl}
\hline
\bf{$T_{1/2}^{2\nu 2\beta}$ (10$^{18}$ yr)}	& \bf{S/B}	& \bf{Experiment}	& \bf{$\beta \beta$ Source} & \bf{$^{100}$Mo exposure}	& \bf{Year [Ref.]}   \\
\hline
7.11$\pm$0.02(stat)$\pm$0.54(syst) & 40	& NEMO-3 & $^{100}$Mo foils			& 7.37 kg$\times$yr			& 2005 \cite{Arnold:2005} \\
6.92$\pm$0.06(stat)$\pm$0.36(syst) & 10	& LUMINEU & Li$_2$$^{100}$MoO$_4$ bol.	& 0.04 kg$\times$yr	& 2017 [This work] \\
\hline
\end{tabular}
\label{tab:DBD_100Mo}
\end{table}

\section{CUPID-0/Mo $2\beta$ EXPERIMENT}
Taking into account the achievements of the LUMINEU project, an extension of the LUMINEU pilot experiment 
to a CUPID demonstrator based on Li$_2$$^{100}$MoO$_4$ scintillating bolometers (CUPID-0/Mo) is in preparation. 
According to the availability of $\sim$7 kg of $^{100}$Mo-enriched 
molybdenum and the cryogenic set-up(s) able to host the LUMINEU-like modules, the CUPID-0/Mo project 
is planned to be realized in two phases:

\begin{itemize}
\item Twenty Li$_2$$^{100}$MoO$_4$ crystals ($\oslash44 \times 45$~mm, $\sim$0.2 kg each; 
2.34 kg of $^{100}$Mo) to be operated as scintillating bolometers in five towers 
inside the EDELWEISS set-up (LSM, France) by the end of 2017.
\item Additional twenty similar-size Li$_2$$^{100}$MoO$_4$-based detectors ready to be operated 
in a complementary set-up\footnote{In principle, the experimental volume of the EDELWEISS set-up 
is able to host also these additional 20 detectors, however a part of the cryostat is going to be dedicated 
to the EDELWEISS low mass WIMPs search program \cite{Arnaud:2017}.}, e.g. CUPID-0 (LNGS, Italy), 
in the middle of 2018.  
\end{itemize}

As one can see in Table \ref{tab:CUPID_Mo}, the sensitivity of the considered CUPID-0/Mo configurations 
would be comparable with the most stringent constraints on the effective Majorana neutrino mass (0.06--0.6 eV) 
derived from the results of the most sensitive $0\nu 2\beta$ experiments \cite{Vergados:2017,Gando:2017}, for which 
a typical exposure is tens--hundreds kg$\times$yr of isotope of interest. It should be noted that 
the sensitivity of the CUPID-0/Mo Phase I remains substantially unaffected even with an order of magnitude 
worse projected background (10$^{-2}$ counts/yr/kg/keV), which is similar to the 0.06$\pm$0.03~counts/yr/kg/keV 
measured in the 2.8--3.6 MeV energy interval by the LUMINEU pilot $2\beta$ 
experiment\footnote{The observed events above 2.65 MeV (see Figure \ref{fig:Background}) 
could be explained by pile-ups of $\gamma$ cascade from $^{208}$Tl decays nearby the detectors and/or 
muon-induced events \cite{Armengaud:2017}. Therefore, we are going to improve the present background 
by removing identified Th contaminated elements and using an available muon veto with a 98\% coverage.}. 
So, in spite of the considerably small scale of the CUPID-0/Mo experiment, this search would be among 
the leading ones in the field. 

\begin{table}[ht]
\caption{The CUPID-0/Mo sensitivity (90\% C.L.) to the $0\nu 2\beta$ decay half-life of $^{100}$Mo 
for different configurations of the experiment. Assumed background is 10$^{-3}$ counts/yr/kg/keV 
in 10 keV window centered at $Q_{\beta \beta}$ of $^{100}$Mo (73\% of decays); 
the efficiency of the pulse shape discrimination is set to be 95\% \cite{Armengaud:2017}.  
The recent calculations of a phase-space factor \cite{Kotila:2012,Stoica:2013}, the nuclear matrix 
elements \cite{Engel:2017,Song:2017}, and an axial-vector coupling constant equal to 1.269 are used 
to estimate the sensitivity to the effective Majorana neutrino mass 
$\left\langle m_{\beta \beta} \right\rangle$.}
\tabcolsep7pt\begin{tabular}{cccc}
\hline
\bf{CUPID-0/Mo configuration}	& \bf{Exposure (kg$\times$yr of $^{100}$Mo)} & \bf{lim$T_{1/2}^{0\nu 2\beta}$ (yr)} & \bf{lim$\left\langle m_{\beta \beta} \right\rangle$ (eV)} \\
\hline
(1) 20$\times$0.5 crystal$\times$yr & 1.2	& 1.3$\times$10$^{24}$ 	& 0.33--0.56 \\
(2) 20$\times$1.5 crystal$\times$yr & 3.5	& 4.0$\times$10$^{24}$ 	& 0.19--0.32 \\
(3) 40$\times$3.0 crystal$\times$yr & 14	& 1.5$\times$10$^{25}$ 	& 0.10--0.17 \\
\hline
\end{tabular}
\label{tab:CUPID_Mo}
\end{table}

\section{CONCLUSIONS}

A production line of large, optical quality, radiopure $^{100}$Mo-enriched Li$_2$$^{100}$MoO$_4$ crystal 
scintillators and their high performance as scintillating bolometers have been established within 
the LUMINEU project. A reasonably high sensitivity to the $0\nu 2\beta$ and the most precise half-life value 
for the $2\nu 2\beta$ decay of $^{100}$Mo (g.s. to g.s. transitions) have been achieved in a pilot 
LUMINEU $2\beta$ experiment over only $\approx$0.1 kg$\times$yr exposure of four 0.2 kg Li$_2$$^{100}$MoO$_4$ 
detectors array operated inside the EDELWEISS set-up in the Modane underground laboratory (France). 
A successful accomplishment of the LUMINEU project triggered its extension to CUPID-0/Mo $2\beta$ 
experiment aiming at operating forty 0.2-kg Li$_2$$^{100}$MoO$_4$ scintillating bolometers. 
The start of data taking with 20 detectors is foreseen in the EDELWEISS set-up by the end of 2017. 
The main CUPID-0/Mo goal is to demonstrate zero-background conditions in the vicinity of the expected 
$^{100}$Mo $0\nu 2\beta$ decay peak and therefore to prove the viability of the LUMINEU technology 
for CUPID project, a ton-scale bolometric $0\nu 2\beta$ experiment.

\section{ACKNOWLEDGMENTS}
This work is a part of LUMINEU program, a project funded by the Agence Nationale de la Recherche (ANR, France). 
The help of the technical staff of the Laboratoire Souterrain de Modane and of the other participant laboratories 
is gratefully acknowledged. We thank the mechanical workshop of CEA/SPEC for its skillful contribution 
to the conception and fabrication of the detectors' holders.

\nocite{*}
\bibliographystyle{aipnum-cp}%
\bibliography{Proc_MEDEX17_Poda_v04}%

\end{document}